\documentclass[conference]{IEEEtran}
\IEEEoverridecommandlockouts

\usepackage{cite}
\usepackage{amsmath,amssymb,amsfonts}
\usepackage{algorithmic}
\usepackage{algorithm}
\usepackage{graphicx}
\usepackage{textcomp}
\usepackage{xcolor}
\usepackage{booktabs} 
\usepackage{multirow}
\usepackage{float}
\usepackage{hyperref}
\usepackage{enumitem}

\setlist[itemize]{topsep=3pt, itemsep=2pt, partopsep=0pt, parsep=0pt}
\setlist[enumerate]{topsep=3pt, itemsep=2pt, partopsep=0pt, parsep=0pt}

\def\BibTeX{{\rm B\kern-.05em{\sc i\kern-.025em b}\kern-.08em
    T\kern-.1667em\lower.7ex\hbox{E}\kern-.125emX}}

\begin{document}

\title{On-Edge Acoustic Continual Anomaly Detection: A Cross-Platform Comparative Study
\thanks{\copyright~2026 IEEE. Personal use of this material is permitted. Permission from IEEE must be obtained for all other uses, in any current or future media, including reprinting/republishing this material for advertising or promotional purposes, creating new collective works, for resale or redistribution to servers or lists, or reuse of any copyrighted component of this work in other works.}
}
\author{
\IEEEauthorblockN{
Rami Zewail\textsuperscript{1,*}, 
Yehia Ehab\textsuperscript{2}, 
Karim  Ali\textsuperscript{3}, 
Ahmed Mahrous\textsuperscript{1}, 
Mohamed Awad\textsuperscript{4}, and 
Bassem Mokhtar\textsuperscript{5}
}
\vspace{1.5mm} 
\IEEEauthorblockA{\textsuperscript{1}\textit{Computer Science \& Engineering Dept., Egypt-Japan Univ. of Science \& Technology}, Alexandria, Egypt}
\IEEEauthorblockA{\textsuperscript{2}\textit{Electronics \& Communications Eng. Dept., Alexandria University}, Alexandria, Egypt}
\IEEEauthorblockA{\textsuperscript{3}\textit{Computer Science \& Engineering Dept., Politecnico di Milano}, Milan, Italy}
\IEEEauthorblockA{\textsuperscript{4}\textit{Statistics and Data Science Dept., Mohamed bin Zayed Univ. of AI (MBZUAI)}, Abu Dhabi, UAE}
\IEEEauthorblockA{\textsuperscript{5}\textit{Computer and Network Engineering Dept., UAE University}, Al Ain, UAE}
\vspace{1.5mm} 
\IEEEauthorblockA{
\textit{Emails:} rami.zewail@ejust.edu.eg , es-ehaborfy@alexu.edu.eg , \\
karimyasser.ali@mail.polimi.it, ahmed.mahrous@ejust.edu.eg,mohamed.awad@mbzuai.ac.ae, 
bassem.mokhtar@uaeu.ac.ae
}
}
\maketitle

\begin{abstract}
Acoustic Anomaly Detection (AAD) serves as a non-contact approach for machine condition monitoring, yet much of the existing research is evaluated offline and rarely addresses embedded hardware constraints or microphone variations under operational drift. This paper presents a comparative evaluation of on-edge continual AAD across two embedded platforms---the ESP32-S3 (with an external I2S digital microphone) and the Arduino Nicla Vision (with an onboard Pulse Density Modulation (PDM) microphone)---using a shared detection model. The model utilizes a lightweight ShallowNet encoder and contrastive prototype learning to map spectrograms into a 32-dimensional embedding space. Both boards run an identical TensorFlow Lite for Microcontrollers (TFLM) encoder, performing on-device prototype learning updated via an exponential moving average. A cloud pipeline orchestrated through Firebase and MQTT enables remote data collection and retraining. Evaluated via a controlled speaker-replay protocol using standardized Detection and Classification of Acoustic Scenes and Events (DCASE) Challenge Task 2 machine sounds, the ESP32-S3 achieved 82.35\% to 95.00\% accuracy across five machine types, while the Nicla Vision achieved 79.41\% to 95.00\%. The two platforms demonstrated comparable performance, confirming that the lightweight prototypical architecture transfers reliably across distinct hardware and sensor tiers.
\end{abstract}


\section{Introduction}
Predictive maintenance relies on monitoring the acoustic signatures of mechanical machinery. Abnormal sounds often indicate wear, imbalance, friction, or blockage. Microphone-based monitoring is non-contact and deployable across multiple machines, making it attractive for scalable industrial condition-monitoring systems.

Most anomalous sound detection (ASD) research is evaluated strictly offline using standardized datasets established by the Detection and Classification of Acoustic Scenes and Events (DCASE) community, such as MIMII \cite{purohit2019mimii}, ToyADMOS \cite{koizumi2019toyadmos}, ToyADMOS2 \cite{harada2021toyadmos2}, and MIMII-DG/DUE \cite{tanabe2021mimiidue, dohi2022mimiidg}. While these benchmarks provide rigorous baselines for algorithmic comparison, offline evaluation does not capture the physical constraints of embedded deployment. In practical field deployments, the edge device must physically capture sound through distinct microphone hardware (such as digital I$^{\text{2}}$S versus onboard PDM Micro-Electromechanical Systems (MEMS) sensors), compute spectral representations locally, run neural inference under strict memory and computational budgets, and communicate decisions to a monitoring layer. Recent TinyML paradigms have enabled lightweight neural architectures on microcontrollers \cite{david2020tflm, warden2019tinyml, banbury2021mlperf}. While lightweight modeling and edge-AI have driven advancements across diverse resource-constrained IoT domains,physical deployment studies in acoustic monitoring remain scarce. Specifically, cross-platform empirical evaluations assessing the same model across differing microcontrollers and physical microphone transducers are largely unexplored.

A fundamental practical question is whether the choice of embedded platform and its associated microphone integration affects detection outcomes under realistic acoustic playback. Integrated sensor architectures like the Arduino Nicla Vision feature onboard PDM sensing, whereas cost-effective modules such as the ESP32-S3 rely on external digital I$^{\text{2}}$S microphones, presenting distinct acoustic transfer characteristics and hardware trade-offs.

This paper examines these aspects through an empirical end-to-end comparison. Our main contributions are: 
\begin{enumerate}
    \item A contrastive prototype learning model with an adaptive percentile threshold and on-device continual adaptation for one-class AAD;
    \item A comprehensive physical acoustic evaluation using machine sound types derived from DCASE benchmark suites (ToyADMOS2 and MIMII/MIMII-DG);
    \item An empirical cross-platform assessment of this shared model across two embedded targets with distinct microphone configurations (ESP32-S3 with external I$^{\text{2}}$S digital microphone vs. Arduino Nicla Vision with integrated PDM MEMS sensor); and
    \item An integrated cloud and IoT telemetry pipeline enabling remote fleet monitoring, data curation, and retraining.
\end{enumerate}

\section{Related Work}
The IEEE DCASE Challenge Task~2 series (from DCASE 2020 through DCASE 2024) has been the central vehicle for advancing unsupervised anomalous sound detection \cite{koizumi2020dcase, kawaguchi2021dcase, nishida2024dcase}. Because industrial anomalies occur unpredictably and damage machinery during collection, the task is formally framed as unsupervised one-class classification: models are trained strictly on normal operational recordings and must flag deviations from the nominal distribution \cite{purohit2019mimii}. Subsequent challenge iterations have highlighted operational domain shifts, such as varying machine operational velocities, load changes, background ambient factory noise, and microphone mismatch \cite{tanabe2021mimiidue, dohi2022mimiidg}.

Early challenge approaches relied on reconstruction error via deep autoencoders or density estimation using Gaussian Mixture Models (GMM) \cite{koizumi2020dcase, xiao2022dcase}. More recent studies have demonstrated the efficacy of metric-based and prototypical meta-learning paradigms, which map acoustic representations into compact latent spaces where normal samples cluster around class prototypes \cite{snell2017prototypical}. Metric learning with contrastive objectives separates heterogeneous operational modes while maintaining compact intra-class variance \cite{khosla2020supervised, koch2015siamese}. 

Deploying these architectures on constrained microcontrollers requires specialized runtimes such as TensorFlow Lite for Microcontrollers (TFLM) \cite{david2020tflm}, coupled with lightweight communication layers like MQTT \cite{banks2014mqtt}. This work unifies prototypical contrastive learning, continual on-device tracking, and TinyML embedded deployment to analyze platform-microphone transferability.

\section{Detection Model}
The identical detection pipeline is executed across both target boards, consisting of a deterministic spectrogram front-end, a ShallowNet projection encoder, and a contrastive prototypical inference engine.

\subsection{Spectrogram Front-End}
Each one-second audio window sampled at $f_{s}=16,000$\,Hz is converted to a log-compressed linear spectrogram. The window is segmented into overlapping frames of length $L_{\text{frame}}=480$ samples (30\,ms) with a frame hop of $L_{\text{step}}=320$ samples (20\,ms):
\begin{equation}
N_{\text{frames}} = \left\lfloor\frac{f_{s} - L_{\text{frame}}}{L_{\text{step}}}\right\rfloor + 1 = \left\lfloor\frac{16000 - 480}{320}\right\rfloor + 1 = 49
\end{equation}

Each frame is zero-padded to $N_{\text{FFT}}=512$, windowed with a symmetric Hamming function, and transformed via Fast Fourier Transform (FFT). The resulting 257 unique magnitude bins are linearly compressed into $N_{\text{buckets}}=43$ frequency buckets (averaging 5 consecutive bins per bucket, discarding the highest bins), followed by a dynamic range compression $\log(1 + x)$. The final $49 \times 43$ matrix is serialized into a 2,107-dimensional feature vector. The system hyperparameter specifications are detailed in Table~\ref{tab:params}.

\subsection{ShallowNet Encoder}
To ensure deterministic execution within strict microcontroller SRAM limits, the encoder network utilizes a compact multilayer perceptron:
\begin{center}
$\text{Input}(2107) \rightarrow \text{Dense}(64, \text{ReLU}) \rightarrow \text{Dense}(32, \text{Linear})$
\end{center}
The architecture comprises approximately 135,000 parameters. Quantized to 8-bit integer precision via TensorFlow Lite post-training quantization, the model requires under 140\,KB of Flash memory and minimal runtime tensor arena space, easily accommodating both embedded platforms.

\subsection{Contrastive Prototype Learning and Continual Adaptation}
Training is conducted exclusively on normal-class machine operational recordings. For each mini-batch or training epoch, a normal operational prototype $p \in \mathbb{R}^{32}$ is established as the empirical centroid:
\begin{equation}
p = \frac{1}{N}\sum_{i=1}^{N} f_{\theta}(x_{i})
\end{equation}
where $f_{\theta}(\cdot)$ represents the quantized neural encoder. The model parameters $\theta$ are trained using a prototype-anchored contrastive Euclidean objective:
\begin{equation}
\mathcal{L} = \frac{1}{B}\sum_{i=1}^{B} \frac{1}{2}\|f_{\theta}(x_{i}) - p\|_{2}^{2}
\end{equation}
The nominal anomaly threshold $\tau$ is determined offline from the 95th percentile of normal operational distances:
\begin{equation}
\tau = \text{Percentile}_{95}\left(\left\{\|f_{\theta}(x_{i}) - p\|_{2}\right\}_{i=1}^{N}\right)
\end{equation}

During on-device boot or machine commissioning, an initial prototype is constructed from the first $N_{p}=30$ captured normal embeddings via recursive running averaging:
\begin{equation}
p^{(k)} = \begin{cases}
e^{(0)}, & k = 0 \\
\frac{e^{(k)} + p^{(k-1)}}{2}, & 1 \le k \le N_{p}
\end{cases}
\end{equation}
To accommodate non-damaging operational drift (e.g., thermal equilibrium shifts, minor load alterations), an on-device continual learning mode updates the active prototype using an exponential moving average with adaptation rate $\alpha = 0.1$:
\begin{equation}
p \leftarrow (1 - \alpha)p + \alpha e
\end{equation}
At inference, Euclidean distance $d(e, p) = \Vert{}e - p\Vert{}_{2}$ is evaluated against $\tau = 0.7$. To prevent transient false alarms, a three-sample unanimous voting window (\texttt{VOTE\_COUNT} = 3) is applied: an anomaly is flagged if and only if three consecutive 1-second frames cross $\tau$ (Algorithm~\ref{alg:prcl}).

\begin{algorithm}[t]
\caption{On-Device Prototypical Representation-Based Continual Learning (PRCL)}
\label{alg:prcl}
\begin{algorithmic}[1]
\REQUIRE Latent embedding $z \in \mathbb{R}^{32}$, learning rate $\alpha$, anomaly threshold $\tau$, max initial samples $N_p = 30$
\ENSURE Anomaly decision $\tilde{y} \in \{\text{Normal}, \text{Anomaly}\}$
\STATE \textbf{Initialization:} Initialize prototype $p \in \mathbb{R}^{32}$ and sample counter $k \leftarrow 0$
\WHILE{System is running}
    \STATE Capture 1-second audio frame via microphone
    \STATE Extract log-linear spectrogram $S \in \mathbb{R}^{49 \times 43}$
    \STATE Obtain latent embedding $e = f_{\theta}(S)$ via TFLM encoder
    \IF{Mode == \textsc{Learning} and $k < N_p$}
        \IF{$k == 0$}
            \STATE $p \leftarrow e$
        \ELSE
            \STATE $p \leftarrow \frac{e + p}{2}$ \COMMENT{Running-average update}
        \ENDIF
        \STATE $k \leftarrow k + 1$
    \ELSIF{Mode == \textsc{Inference}}
        \STATE Compute Euclidean distance: $d(e, p) = \vert{}\vert{}e - p\vert{}\vert{}_{2}$
        \IF{$d(e, p) > \tau$}
            \STATE $\tilde{y} \leftarrow \text{Anomaly}$
        \ELSE
            \STATE $\tilde{y} \leftarrow \text{Normal}$
        \ENDIF
        \IF{Mode == \textsc{Continual\_Learning} and $\tilde{y} == \text{Normal}$}
            \STATE $p \leftarrow (1 - \alpha)p + \alpha \, e$ \COMMENT{Exponential moving average update}
        \ENDIF
        \STATE Apply 3-sample unanimous voting window
    \ENDIF
\ENDWHILE
\end{algorithmic}
\end{algorithm}

\begin{table}[t]
\caption{Shared System Hyperparameters and Configuration}
\label{tab:params}
\centering
\begin{tabular}{llc}
\toprule
\textbf{Parameter} & \textbf{Symbol / Identifier} & \textbf{Value} \\
\midrule
Audio Sample Rate & \texttt{SAMPLE\_RATE} & 16,000\,Hz \\
FFT Frame Length & \texttt{FRAME\_LEN} & 480 (30\,ms) \\
FFT Frame Hop & \texttt{FRAME\_STEP} & 320 (20\,ms) \\
Number of Temporal Frames & \texttt{NUM\_FRAMES} & 49 \\
FFT Size & \texttt{FFT\_SIZE} & 512 \\
Frequency Buckets & \texttt{NUM\_BUCKETS} & 43 \\
Serialized Input Dimension & -- & 2,107 \\
Encoder Hidden Layer & \texttt{HIDDEN\_DIM} & 64 (ReLU) \\
Encoder Output Dimension & \texttt{OUTPUT\_LEN} & 32 (Linear) \\
Anomaly Detection Threshold & \texttt{SIMILARITY\_THRESHOLD} & 0.7 \\
Temporal Voting Window & \texttt{VOTE\_COUNT} & 3 (unanimous) \\
Calibration Prototype Count & \texttt{PROTOTYPE\_COUNT} & 30 \\
Continual Adaptation Rate & \texttt{LEARN\_RATE} ($\alpha$) & 0.1 \\
\bottomrule
\end{tabular}
\end{table}

\section{Cloud Pipeline}
To support remote lifecycle management, an IoT cloud architecture interfaces with edge nodes. Firebase Firestore maintains operational states, detection thresholds, and control parameters. Firebase Realtime Database stores telemetry logs and raw spectrograms. Model artifacts and quantization profiles are stored in Supabase Storage, while a centralized dashboard hosted via Hugging Face Spaces provides an interface for model retraining. The physical microcontrollers communicate bi-directionally with a Mosquitto MQTT broker over secure WebSockets and TLS, publishing anomaly states and receiving calibration updates.

\section{Embedded Deployment}
The edge framework was implemented identically across two target hardware environments:
\begin{itemize}
    \item \textbf{ESP32-S3:} Equipped with a dual-core Xtensa 32-bit LX7 processor (240\,MHz, 512\,KB SRAM, 8\,MB PSRAM, 16\,MB Flash). It interfaces with an external MIC V829 MEMS microphone over an I$^{\text{2}}$S digital bus configured in master receive mode (16-bit, 16\,kHz, left channel) using GPIO~6 (WS), GPIO~5 (SCK), and GPIO~7 (SD).
    \item \textbf{Arduino Nicla Vision:} Equipped with a dual-core STM32H747AII6 MCU (Cortex-M7 at 480\,MHz and Cortex-M4 at 240\,MHz, 1\,MB SRAM, 2\,MB Flash). It utilizes an onboard STMicroelectronics MP34DT05 PDM digital microphone, captured using hardware decimation and PDM-to-PCM conversion.
\end{itemize}

\begin{figure}[t]
\centering
\includegraphics[width=0.88\columnwidth]{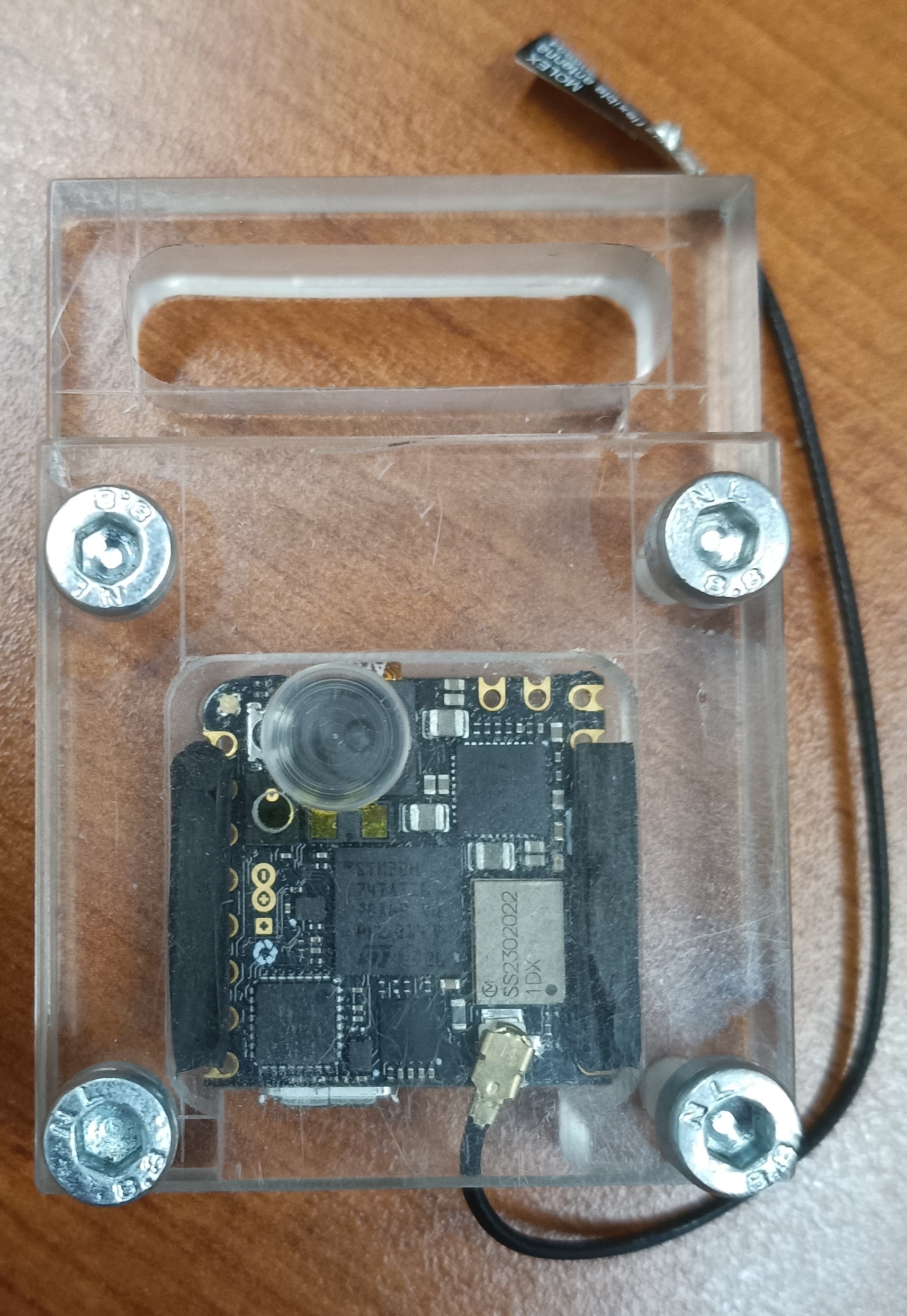}
\caption{Encased Arduino Nicla Vision deployment platform featuring external antenna connectivity and secure acrylic mounting for physical playback evaluation.}
\label{fig:hardware_setup}
\end{figure}

\section{Experimental Setup}

\subsection{Speaker-Replay Protocol and Microphone Evaluation}
Physical evaluation was performed using a standardized acoustic speaker-replay protocol. Audio recordings were emitted via the calibrated internal sound system of an Apple MacBook Pro (M3 Pro) located within an enclosed room (ambient background $\approx 38$\,dBA). The target embedded devices were placed at a fixed distance of 30\,cm on-axis from the sound source. The microphones (external I$^{\text{2}}$S on ESP32-S3 vs. onboard PDM on Nicla Vision) captured the replayed acoustic waveforms, subject to ambient room reverberation, multipath reflections, and loudspeaker transfer functions. This replicates the real-world operational reality where factory monitoring devices capture sound across distinct acoustic paths and sensor transductions.

\begin{figure}[t]
\centering
\includegraphics[width=\columnwidth]{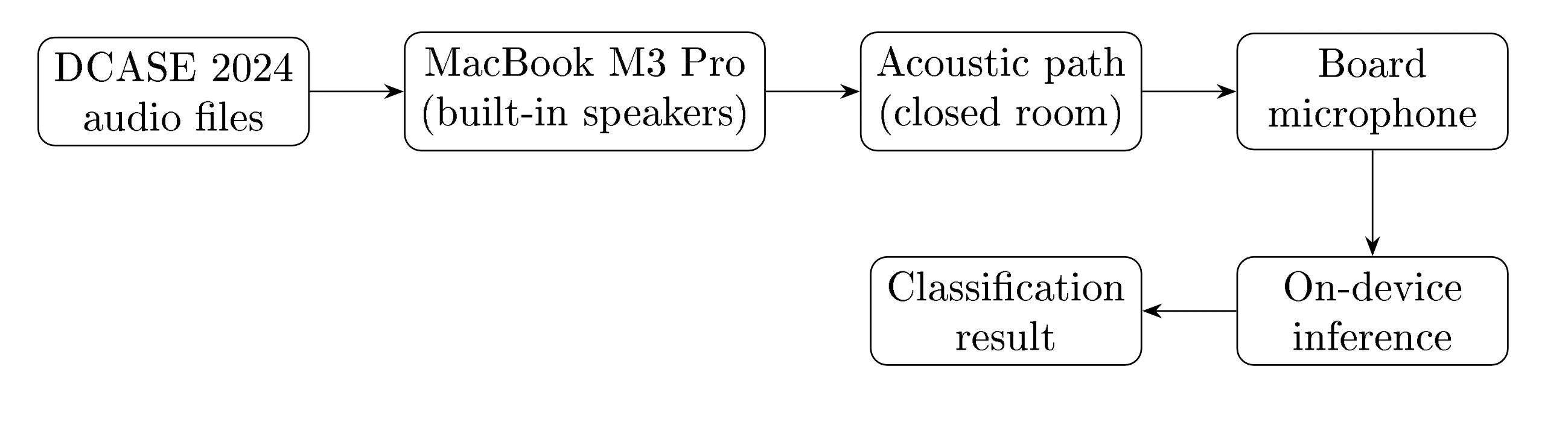}
\caption{Controlled physical speaker-replay experimental pipeline. Calibrated DCASE benchmark machine recordings are emitted via a reference loudspeaker in a closed room and captured by edge microcontrollers running local TFLM inference.}
\label{fig:replay}
\end{figure}

\subsection{Benchmark Dataset Composition and Machine Types}
The evaluation audio was drawn from the internationally recognized DCASE Challenge Task~2 benchmark suites \cite{koizumi2020dcase, kawaguchi2021dcase, nishida2024dcase}, which merge two cornerstone industrial datasets:
\begin{enumerate}
    \item \textbf{ToyADMOS / ToyADMOS2 Dataset \cite{koizumi2019toyadmos, harada2021toyadmos2}:} Provides operating acoustic recordings of miniature machines under controlled, reproducible anomalous conditions. The evaluated classes include:
    \begin{itemize}
        \item \textbf{ToyCar:} Toy miniature vehicle sounds exhibiting operational anomalies such as gear wear, shaft friction, and motor irregularities under varying operational speeds \cite{harada2021toyadmos2}.
        \item \textbf{Toy Train:} Model railroad vehicles operating on rails, exhibiting track friction, mechanical clatter, wheel blockages, and motor defects \cite{harada2021toyadmos2}.
    \end{itemize}
    \item \textbf{MIMII / MIMII-DUE / MIMII-DG Datasets \cite{purohit2019mimii, tanabe2021mimiidue, dohi2022mimiidg}:} Provide real-world industrial machine operating sounds captured under genuine industrial noise conditions (such as background factory floor hum, environmental wind, and SNR variations ranging from $-12$\,dB to $+12$\,dB ). The evaluated classes include:
    \begin{itemize}
        \item \textbf{Fan:} Industrial ventilation fans operating with normal airflow vs. structural anomalies, including wing unbalance, blade deformation, clogging, and bearing friction \cite{purohit2019mimii}.
        \item \textbf{Gearbox:} Industrial reduction gearboxes running under load, with anomalous signatures arising from missing gear teeth, surface pitting, and mechanical backlash \cite{tanabe2021mimiidue, dohi2022mimiidg}.
        \item \textbf{Slider (Slide Rail):} Linear slider stages driven by ball screws and belts, with anomalies reflecting rail degradation, loose belts, track debris, and insufficient lubrication \cite{purohit2019mimii}.
    \end{itemize}
\end{enumerate}

\begin{table}[t]
\caption{Characteristics of the Evaluated DCASE Machine Datasets}
\label{tab:dcase_composition}
\centering
\begin{tabular}{lp{2.6cm}cc}
\toprule
\textbf{Machine Type} & \textbf{Benchmark Source Dataset} & \textbf{Normal ($n$)} & \textbf{Abnormal ($n$)} \\
\midrule
ToyCar & ToyADMOS2 \cite{harada2021toyadmos2} & 17 & 17 \\
Toy Train & ToyADMOS2 \cite{harada2021toyadmos2} & 17 & 17 \\
Fan & MIMII / MIMII-DG \cite{purohit2019mimii, dohi2022mimiidg} & 17 & 17 \\
Gearbox & MIMII-DUE / MIMII-DG \cite{tanabe2021mimiidue, dohi2022mimiidg} & 20 & 20 \\
Slider & MIMII / MIMII-DG \cite{purohit2019mimii, dohi2022mimiidg} & 20 & 20 \\
\midrule
\textbf{Total} & \textbf{DCASE Task 2 Suite} & \textbf{91} & \textbf{91} \\
\bottomrule
\end{tabular}
\end{table}

Table~\ref{tab:dcase_composition} summarizes the test split characteristics across the five machines. In total, 182 individual 10-second machine audio files (91 normal, 91 abnormal) were evaluated via on-device physical replay. Standard classification performance metrics are reported, treating the abnormal (anomalous) class as the positive class. Let TP, TN, FP, and FN denote true positives, true negatives, false positives, and false negatives, respectively:
\begin{equation}
\text{Accuracy} = \frac{\text{TP} + \text{TN}}{\text{TP} + \text{TN} + \text{FP} + \text{FN}}
\end{equation}
\begin{equation}
\text{Precision} = \frac{\text{TP}}{\text{TP} + \text{FP}}, \quad \text{Recall} = \frac{\text{TP}}{\text{TP} + \text{FN}}
\end{equation}
\begin{equation}
\text{F1-Score} = \frac{2 \cdot \text{Precision} \cdot \text{Recall}}{\text{Precision} + \text{Recall}}
\end{equation}

\section{Results and Discussion}

\subsection{ESP32-S3 Performance}
The empirical classification metrics recorded by the ESP32-S3 paired with its external I$^{\text{2}}$S digital microphone are presented in Table~\ref{tab:esp32}. The overall accuracy spanned from 82.35\% on ToyCar and Fan to 95.00\% on Slider. The system exhibited balanced precision and recall across all five machine categories, demonstrating the ability of the 32-dimensional embedding space to distinguish acoustic shifts under physical room acoustic replay.

\begin{table}[t]
\caption{ESP32-S3 Classification Metrics (Speaker-Replay Protocol)}
\label{tab:esp32}
\centering
\setlength{\tabcolsep}{4pt}
\begin{tabular}{lccccc}
\toprule
\textbf{Machine} & \textbf{$n$} & \textbf{Acc. (\%)} & \textbf{Prec. (\%)} & \textbf{Rec. (\%)} & \textbf{F1 (\%)} \\
\midrule
ToyCar & 34 & 82.35 & 82.35 & 82.35 & 82.35 \\
Toy Train & 34 & 91.18 & 88.89 & 94.12 & 91.43 \\
Fan & 34 & 82.35 & 82.35 & 82.35 & 82.35 \\
Gearbox & 40 & 87.50 & 85.71 & 90.00 & 87.80 \\
Slider & 40 & 95.00 & 95.00 & 95.00 & 95.00 \\
\bottomrule
\end{tabular}
\end{table}

\subsection{Nicla Vision Performance}
Table~\ref{tab:nicla} reports the corresponding metrics registered by the Arduino Nicla Vision using its integrated STMicroelectronics PDM MEMS microphone. Accuracy ranged between 79.41\% (ToyCar) and 95.00\% (Slider). Notably, the Nicla Vision attained a 100.00\% recall on the Toy Train machine category (detecting all 17 anomalies without a single false negative), while achieving 88.24\% accuracy on the industrial Fan class. On both Gearbox and Slider, the Nicla Vision matched the ESP32-S3 total classification accuracy exactly.

\begin{table}[t]
\caption{Arduino Nicla Vision Classification Metrics (Speaker-Replay Protocol)}
\label{tab:nicla}
\centering
\setlength{\tabcolsep}{4pt}
\begin{tabular}{lccccc}
\toprule
\textbf{Machine} & \textbf{$n$} & \textbf{Acc. (\%)} & \textbf{Prec. (\%)} & \textbf{Rec. (\%)} & \textbf{F1 (\%)} \\
\midrule
ToyCar & 34 & 79.41 & 77.78 & 82.35 & 80.00 \\
Toy Train & 34 & 94.12 & 89.47 & 100.00 & 94.44 \\
Fan & 34 & 88.24 & 84.21 & 94.12 & 88.89 \\
Gearbox & 40 & 87.50 & 89.47 & 85.00 & 87.18 \\
Slider & 40 & 95.00 & 95.00 & 95.00 & 95.00 \\
\bottomrule
\end{tabular}
\end{table}

\subsection{Cross-Platform and Transducer Comparison}
\begin{figure}[t]
\centering
\includegraphics[width=\columnwidth]{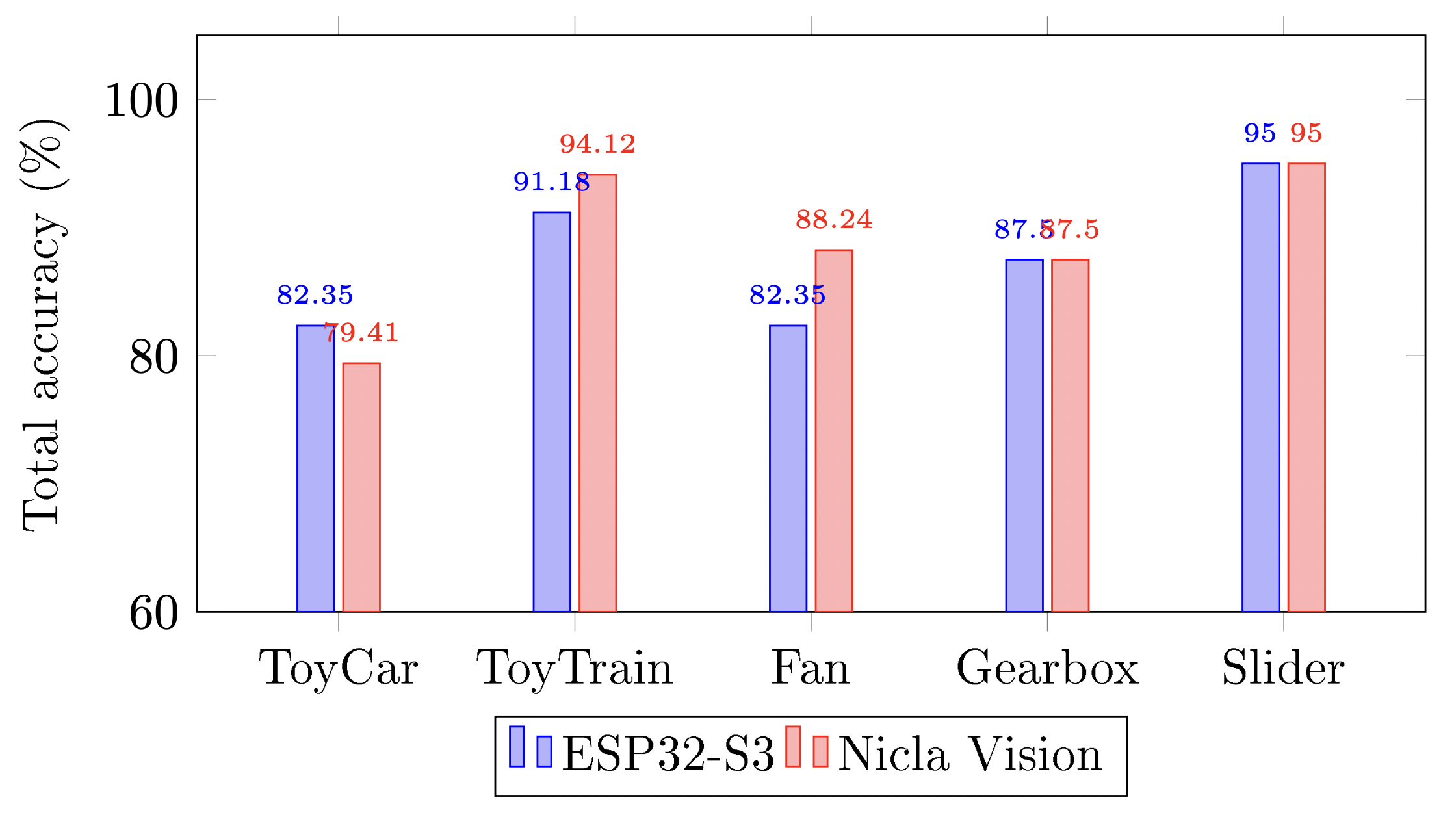}
\caption{Cross-platform classification accuracy by DCASE machine type}
\label{fig:results}
\end{figure}

A direct comparison between the two platforms reveals remarkable consistency, confirming that the prototypical contrastive formulation provides strong acoustic feature invariance across physical transducer hardware. As illustrated in Fig.~\ref{fig:results}, the ESP32-S3 held an advantage on the ToyCar dataset (+2.94 percentage points), whereas the Nicla Vision demonstrated superior detection accuracy on Toy Train (+2.94 percentage points) and Fan (+5.89 percentage points). On the Gearbox and Slider classes, both platforms achieved identical overall accuracy scores of 87.50\% and 95.00\%, respectively.

An analysis of the confusion counts summarized in Table~\ref{tab:fp_fn} illuminates subtle transducer characteristics. On Gearbox, the two platforms exhibited complementary error tradeoffs: the external I$^{\text{2}}$S sensor on the ESP32-S3 achieved higher recall (90.00\% vs. 85.00\%, incurring only 2 false negatives), while the onboard PDM microphone on the Nicla Vision yielded higher precision (89.47\% vs. 85.71\%, incurring fewer false positives). On the linear Slider stage, both boards achieved identical, highly balanced performance with only a single false positive and false negative out of 40 trials.

\begin{table}[t]
\caption{Detailed Error Distribution: False Positive (FP) and False Negative (FN) Counts}
\label{tab:fp_fn}
\centering
\begin{tabular}{lcccc}
\toprule
\multirow{2}{*}{\textbf{Machine Type}} & \multicolumn{2}{c}{\textbf{ESP32-S3 (I$^{\text{2}}$S)}} & \multicolumn{2}{c}{\textbf{Nicla Vision (PDM)}} \\
\cmidrule(r){2-3} \cmidrule(l){4-5}
& \textbf{FN} & \textbf{FP} & \textbf{FN} & \textbf{FP} \\
\midrule
ToyCar & 3 & 3 & 3 & 4 \\
Toy Train & 1 & 2 & 0 & 2 \\
Fan & 3 & 3 & 1 & 3 \\
Gearbox & 2 & 3 & 3 & 2 \\
Slider & 1 & 1 & 1 & 1 \\
\midrule
\textbf{Total Errors} & \textbf{10} & \textbf{12} & \textbf{8} & \textbf{12} \\
\bottomrule
\end{tabular}
\end{table}

\begin{figure}[t]
\centering
\includegraphics[width=0.75\columnwidth]{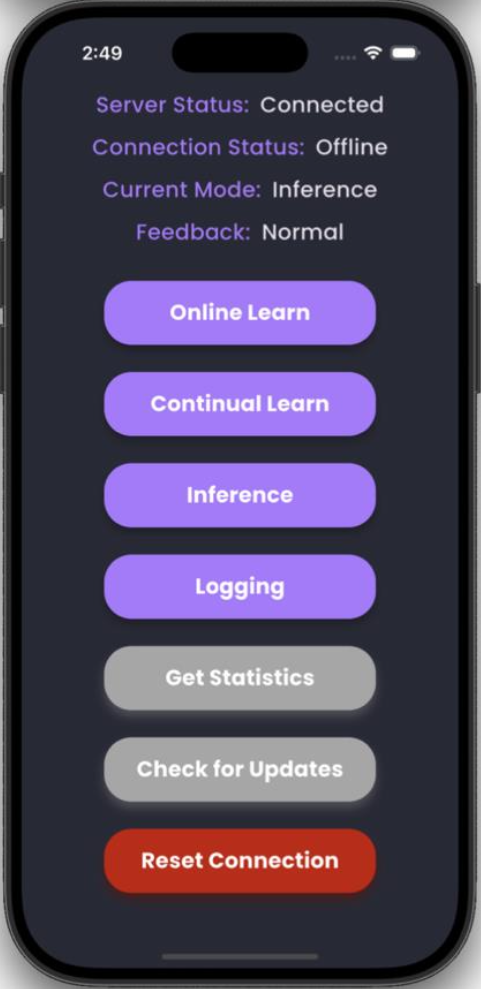}
\caption{Mobile monitoring dashboard receiving real-time anomalous sound notifications, embedding distances, and health status published via MQTT.}
\label{fig:mobile_app}
\end{figure}

These findings indicate that despite differences in microphone frequency response, onboard analog front-end filtering, and digital decimation pipelines, the shared 8-bit quantized ShallowNet model combined with log-compressed linear binning maintains robust anomaly detection capabilities without requiring board-specific retraining.

\section{Limitations and Future Work}
While the speaker-replay protocol enables rigorous multi-board benchmarking under controlled acoustic conditions, physical sound reproduction through loudspeakers introduces specific frequency coloration and lacks the structural casing vibrations encountered during direct machinery attachment. Additionally, the evaluation sets of 34 to 40 samples per class, while sufficient for comparative feasibility, represent small sample sizes. Furthermore, prototype embeddings currently reside in volatile SRAM during operation and must be re-calibrated upon reboot. Future efforts will integrate non-volatile prototype persistence in flash memory, evaluate multichannel acoustic arrays, and extend on-device continual adaptation to distributed federated architectures across operational industrial machinery.

\section{Conclusion}
This paper presented an empirical investigation of on-edge acoustic anomaly detection across two embedded platforms---the ESP32-S3 and the Arduino Nicla Vision---evaluating the transferability of a shared prototypical contrastive learning model under distinct microphone integrations. Benchmarked across five machine types curated from the DCASE Challenge Task~2 suites (ToyADMOS2 and MIMII/MIMII-DG), the ESP32-S3 achieved 82.35\% to 95.00\% accuracy and the Nicla Vision achieved 79.41\% to 95.00\% accuracy. Both platforms tied on two industrial machine classes (87.50\% on Gearbox and 95.00\% on Slider), demonstrating that the lightweight prototypical architecture operates consistently across distinct microcontroller architectures and MEMS microphone configurations.

\section*{Acknowledgment}
This work was supported by the Information Technology Industry Development Agency (ITIDA) - Information Technology Academia Collaboration (ITAC) program under grant number CFP247. The author(s) also acknowledge the use of Paperpal for grammar correction and stylistic refinement during the preparation of this manuscript, and assume full responsibility for the final publication.

\bibliographystyle{IEEEtran}
\bibliography{references}

\end{document}